\documentclass[12pt]{article}
\usepackage{amsmath,amssymb,amsfonts}
\usepackage[paper=letterpaper,margin=1.0in]{geometry}
\usepackage{graphicx}

\newcommand{\be}{\begin{equation}}
\newcommand{\ee}{\end{equation}}
\newcommand{\bea}{\begin{eqnarray}}
\newcommand{\eea}{\end{eqnarray}}
\newcommand{\ba}{\begin{array}}
\newcommand{\ea}{\end{array}}
\newcommand{\ben}{\begin{enumerate}}
\newcommand{\een}{\end{enumerate}}
\newcommand{\bi}{\begin{itemize}}
\newcommand{\ei}{\end{itemize}}
\newcommand{\bc}{\begin{center}}
\newcommand{\ec}{\end{center}}
\newcommand{\bfig}{\begin{figure}}
\newcommand{\efig}{\end{figure}}
\newcommand{\bq}{\begin{quotation}}
\newcommand{\eq}{\end{quotation}}
\newcommand{\bt}{\begin{table}}
\newcommand{\et}{\end{table}}
\newcommand{\btab}{\begin{tabular}}
\newcommand{\etab}{\end{tabular}}
\newcommand{\bs}{\begin{slide}}
\newcommand{\es}{\end{slide}}

\newcommand{\eref}[1]{(\ref{#1})}

\newcommand{\AdS}[1]{{\rm AdS}_{#1}}
\newcommand{\pa}{\partial}

\begin{document}

{\footnotesize
\rightline{QMUL-PH-09-30}
\rightline{VPI-IPNAS-09-14}
}

\bc

\vskip 1.0cm
\centerline{\Large \bf String Theory and Turbulence}
\vskip 0.5cm
% \centerline{[Version of \today]}
\vskip 1.0cm

\renewcommand{\thefootnote}{\fnsymbol{footnote}}

\centerline{{\bf
Vishnu Jejjala${}^{1}$\footnote{\tt v.jejjala@qmul.ac.uk},
Djordje Minic${}^{2}$\footnote{\tt dminic@vt.edu},
Y.\ Jack Ng${}^{3}$\footnote{\tt yjng@physics.unc.edu}, and
Chia-Hsiung Tze${}^{2}$\footnote{\tt kahong@vt.edu}}}

\vskip 0.5cm

{\it
${}^1$Centre for Research in String Theory \\ Department of Physics, Queen Mary, University of London \\ Mile End Road, London E1 4NS, U.K. \\
${}$ \\
${}^2$Institute for Particle, Nuclear and Astronomical Sciences \\ Department of Physics, Virginia Tech \\ Blacksburg, VA 24061, U.S.A. \\
${}$ \\
${}^3$Institute of Field Physics \\ Department of Physics and Astronomy, University of North Carolina \\ Chapel Hill, NC 27599, U.S.A.
}

\ec

\vskip 1.0cm

\begin{abstract}
We propose a string theory of turbulence that explains the Kolmogorov scaling in $3+1$ dimensions and the Kraichnan and Kolmogorov scalings in $2+1$ dimensions.
This string theory of turbulence should be understood in light of the AdS/CFT dictionary.
Our argument is crucially based on the use of Migdal's loop variables and the self-consistent solutions of Migdal's loop equations for turbulence.
In particular, there is an area law for turbulence in $2+1$ dimensions related to the Kraichnan scaling.
\end{abstract}

\newpage

Turbulence is one of the great unsolved problems of physics~\cite{review}.
It is as well an exceptional proving ground for ideas about strong coupling, strong correlations, non-linearity, complexity, and far from equilibrium physics.
The remarkable fact about fully developed turbulence in {\em three} spatial dimensions is that it obeys the well-known Kolmogorov scaling law~\cite{kol}.
By contrast, in addition to the Kolmogorov scaling, in {\em two} spatial dimensions fully developed turbulence exhibits other scaling behavior, in particular Kraichnan scaling~\cite{kr}.

We have recently argued that there are deep similarities between quantum gravity and turbulence~\cite{previous}.
The connection between these seemingly disparate fields is provided by the r\^ole of the diffeomorphism symmetry in classical gravity and the volume preserving diffeomorphisms of classical fluid dynamics.
In computing correlators of the velocity field, one is led to examine the statistical and Euclidean quantum field theoretic descriptions of turbulence.
By utilizing the metrical properties of sound propagation in fluids, we have argued that the Kolmogorov scaling in $3+1$ dimensions can be derived from quantum gravity, in particular, the features of a holographic spacetime foam.
In these investigations, we have uncovered a friction between holography and Kraichnan scaling in $2+1$ dimensions and suggested that this may relate to strong coupling dynamics in the ultraviolet.

In this paper, we sharpen the intuition about the relation between turbulence and quantum gravity and propose a very specific dictionary between string theory and turbulence.
We argue that the relation between the Kolmogorov and Kraichnan scalings is precisely the same as the one between the string and membrane theories.
We also argue that the AdS/CFT correspondence finds its natural ``turbulent'' realization in this context.
This opens up the possibility, analogous to the proposal of the QCD string, for mapping the solutions of sigma models in particular backgrounds to the various statistical distributions associated with turbulent flows.

In what follows, we retrace very closely the seminal discussion of turbulence in terms of loop equations proposed by Migdal~\cite{migdal}.
The basic equation of turbulent fluid dynamics is the Navier--Stokes equation
\be
\rho ( \pa_t v_i + v_j\, \pa_j v_i ) = -\pa_i p + \nu\, \pa_j^2 v_i ~,
\ee
with $\pa_i v_i = 0$, in the limit of infinite Reynolds number, which formally amounts to setting the viscosity $\nu$ to zero.
(The velocity field of the flow is $v_i$, $p$ is pressure, and $\rho$ is the fluid density.)
The object is to compute the generating functional of the velocity correlators:
\be
Z(j) = \left\langle \exp\left(-\int d^3x\ j_i(x) v_i(x)\right) \right\rangle ~.
\ee
Our proposal will be that this generating functional can be determined through the AdS/CFT correspondence in the Kraichnan regime.
Furthermore, the Kolmogorov and Kraichnan regimes will be related in a way suggested by the AdS/CFT correspondence in the context of the gauge theory/membrane theory transition in $2+1$ dimensions.

To motivate this proposal, we shall follow Migdal~\cite{migdal} and rewrite the Navier--Stokes equation in terms of loop variables.
Consider the turbulent loop functional
\be
W(C) \sim \exp\left(-\frac{1}{\nu}\int_C dx_i\ v_i \right) ~.
\ee
The viscosity has the dimensions of length squared over the unit of time.
The loop functional can be rewritten using the vorticity field
\be
\omega_{ij} \equiv \pa_i v_j - \pa_j v_i
\ee
using Stokes' theorem
\be
W(C) \sim \exp\left(-\frac{1}{\nu}\int_S d\sigma_{ij}\ \omega_{ij}\right) ~,
\ee
where $\pa S \equiv C$.
The velocity acts as an effective vector potential, and the vorticity is its curl, and thus an effective magnetic field.

As Migdal explains~\cite{migdal}, the Navier--Stokes equations can be formulated as an effective Schr\"odinger equation involving the loop functional $W(C)$
\be
i\, \nu\, \pa_t W(C) = H_C\, W(C) ~,
\label{eq:loop}
\ee
where the loop equation Hamiltonian $H_C$ is
\be
H_C \equiv \nu^2 \int_C dx_j\ \left( i\, \pa_k \frac{\delta}{\delta \sigma_{kj}(x)} + \int d^3y\ \frac{y_l - x_l}{4\pi\, |y-x|^3} \frac{\delta^2}{\delta \sigma_{kj}(x)\delta \sigma_{kl}(y)} \right) ~.
\ee
Note that the viscosity plays the role of $\hbar$ in these turbulent loop equations.
In discussing fully developed turbulence, one always takes the $\hbar \to 0$ or zero viscosity limit.
As viscosity is a dimensionful quantity, this means simply that the dissipative term is much smaller than the non-linear convective derivative.
The perturbation expansion is organized in powers of $\nu$.
As pointed out in~\cite{migdal}, the loop Hamiltonian is not Hermitian due to dissipation.
Moreover, it is non-local.
In our discussion of the turbulent Kolmogorov and Kraichnan distributions, we will be interested in the $\nu\to 0$ limit, {\it i.e.}, a semiclassical asymptotics.
The theory is unitary in this limit.

The Kolmogorov scaling~\cite{kol} follows from the assumption of constant energy flux.
We have
\be
\frac{v^2}{t} \sim \varepsilon ~,
\ee
where the single length scale $\ell$ is given as $\ell \sim v\cdot t$.
This implies that
\be
v \sim (\varepsilon\, \ell)^{1/3} ~.
\ee
Following Migdal, we insert this as a self-consistent ansatz for the loop functional (the area being naturally $A \sim \ell^2$):
\be
W_{\rm Kol} \sim \exp\left(-\frac{\alpha}{\nu}\, \varepsilon^{1/3}\, A^{2/3}\right) ~,
\label{eq:kol}
\ee
which constitutes Migdal's central observation~\cite{migdal}.
(The $\alpha$ in the exponential is an undetermined real constant.)

This behavior should be contrasted with the area law associated with the confining phase of the pure Yang--Mills theory.
From the area law of the turbulent loop, one deduces that the energy spectrum follows the $k^{-5/3}$ law, or in real space, to the experimentally observed two-point function
\be
\langle v^i(\ell) v^j(0)\rangle \sim (\varepsilon\, \ell)^{2/3} \delta^{ij} ~.
\ee
(The energy scaling $E(k) \sim k^{-n}$ is determined as the one-dimensional Fourier transform of the scaling of the two-point function for the velocity field $\langle v(\ell)^2 \rangle \sim \ell^{n-1}$~\cite{review}.)

That we have a self-consistent ansatz for the solution of the loop equation~\eref{eq:loop} in the Kolmogorov regime is seen as follows.
Migdal takes a WKB ansatz for the loop functional~\cite{migdal}:
\be
W(C) = \exp\left(- \frac{1}{\nu} S(C) \right) ~.
\ee
The effective equation for $S(C)$, derived from the loop equation for $W(C)$, is a non-linear Hamilton--Jacobi equation for which Migdal finds self-consistent solutions.
The loop equation for $W(C)$ provides a rewriting of the Navier--Stokes equations in terms of the appropriate collective variables.
By looking at the original Navier--Stokes equations, one immediately notices that for the case of zero viscosity $\nu \to 0$ there exists a self-consistent scaling law
\be
v \sim \ell^{\gamma}
\ee
because the time derivative of velocity and the convective derivative have the same scaling dimension.
This scaling law is naturally violated for finite viscosity, because the dissipative term $\nabla^2 \vec{v}$ has a different scaling dimension.

In general, the scaling coefficient $\gamma$ is not determined, but it immediately leads to the scaling law for $S(C)$:
\be
S(C) \sim \ell^{\gamma+1} \sim t^{\frac{\gamma+1}{1-\gamma}} \equiv t^{2\kappa-1} ~.
\ee
For the case of Kolmogorov scaling $\gamma=1/3$, or, following Migdal's notation, $\kappa \equiv \frac{1}{1-\gamma} = 3/2$, and yields $S(C) \sim t^2$.
The requirement that the three-point function
\be
\langle v_a v_b \partial_a v_b \rangle \sim {\rm constant} ~,
\label{eq:3Kol}
\ee
given the above scaling for the velocity, is equivalent to the Kolmogorov's $\gamma=1/3$.
One can easily see, by applying chain rule, that the constancy of the three-point function amounts to $v^2/t \sim {\rm constant}$, which is the original requirement imposed by Kolmogorov.

We can now extrapolate the observation of Migdal to Kraichnan scaling in $2+1$ dimensions.
In the two-dimensional case there is a second conserved quantity, the {\em enstrophy}
\be
\Omega = \int d^2x\ \omega^2 ~,
\ee
where $\omega$ is the vorticity vector $\vec{\omega} \equiv \nabla \times \vec{v}$ introduced above.\footnote{
In $2+1$ dimensions, the quantities $\Omega_n = \int d^2x\ \omega^{2n}$ are conserved as well.
This infinite tower of conserved currents suggests the existence of an integrable structure both at the classical and the quantum level.}
According to Kraichnan, the constant flux of enstrophy gives
\be
\frac{\omega^2}{t} \sim {\rm constant}
\ee
and implies that the statistical velocity field scales as
\be
v \sim \frac{\ell}{t_0}
\ee
where $t_0$ is the characteristic constant.
This leads to the $k^{-3}$ scaling of the energy in momentum space.
Thus in $2+1$ dimensions we have both the energy (Kolmogorov) and enstrophy (Kraichnan) cascades.
Given the Kraichnan scaling, the turbulent loop goes as
\be
W_{\rm Kr} \sim \exp\left(-\frac{\alpha}{\nu\, t_0}\, A\right) ~.
\label{eq:kr}
\ee

This behavior can be understood in the following sense.
For Kraichnan scaling, the exponent $\gamma$ is fixed by demanding the constancy of the three-point function
\be
\langle v_a \omega_b \partial_a \omega_b \rangle \sim {\rm constant} ~,
\label{eq:3Kr}
\ee
which is equivalent, again by using the chain rule, to saying that $\omega^2/t \sim {\rm constant}$, or $v\sim \ell$ ($\gamma=1$).
Note that rephrasing the Kolmogorov and the Kraichnan behaviors in this way allows for the interpretation of both scalings in terms of a quantum field theoretic anomaly~\cite{polyakov, gawedzki}.

Even within this quantum field theoretic setting a question remains:
why are the Kraichnan and Kolmogorov scalings true?
Our main point is that by looking at Migdal's loop functional the natural geometric area law leads to the Kraichnan scaling, and furthermore to the Kolmogorov scaling by invoking the relation between strings and membranes.
The area law can be naturally interpreted from the point of view of the AdS/CFT correspondence~\cite{adscft} thus opening a connection between string theory and turbulence.

The scaling of the turbulent loop in~\eref{eq:kr} gives the same area law as in the case of the confining Yang--Mills theory.
We claim that this is indicative of some effective Nambu--Goto area action of string theory.
We first notice that the area law is natural for the large Wilson loops in two spatial dimensions, in the case of planar geometry.
Noting that volume $V\sim \ell^3$, the area law for the Kraichnan scaling can be rewritten in terms of the volume of the spacetime filling membrane in $2+1$ dimensions.
This leads to a $2/3$ power:
\be
\exp(-f\, A) = \exp(-f\, V^{2/3}) ~.
\label{eq:23}
\ee
Crucially, the volume in the previous relation is the worldvolume of a surface:
the turbulent loop, which is a string, thickens into a membrane.
This is motivated by the energy cascade in which the turbulent loop breaks up into many smaller loops, whose worldsheets give rise to a membrane picture, as we argue below.
Importantly, the $2/3$ exponent is common to both the Kolmogorov and Kraichnan scalings.
(The prefactor $f$ in~\eref{eq:23} absorbs the dimensionful constants.)
If one allows for a membrane/string transition, one might expect the Kolmogorov cascade to emerge in the case when the membrane volume is replaced by the area of the string worldsheet:
\be
V \to A \qquad \Longrightarrow \qquad \exp(-f\, V^{2/3}) \to \exp(-f\, A^{2/3}) ~.
\ee
This would indicate that {\em both the Kraichnan and the Kolmogorov scalings should be understood from a unified point of view in $2+1$ dimensions}.

Let us recapitulate.
The first result for $W_{\rm Kol}$ in~\eref{eq:kol} should be compared to the usual area law of the Yang--Mills theory.
We might call the Kolmogorov result in this case a turbulent deformation of the area law, so that
\be
\exp(-f\, A) \Longrightarrow \exp(-f\, A^{2/3}) ~.
\ee
For the three-dimensional Yang--Mills theory, which is not conformal as the coupling is dimensionful, we would have the same deformation.
But the AdS/CFT correspondence instructs us that in three dimensions we can have another theory --- the theory of interacting membranes.
For the membrane theory, dual to M-theory on $\AdS{4}$, we should have a volume law associated with surface operators.
This point of view has been advanced in the classic papers on the evaluation of Wilson loops in AdS/CFT~\cite{maldacena}.
Thus we can view the Kraichnan scaling as a turbulent deformation of the volume law
\be
\exp(-f\, V) \Longrightarrow \exp(-f\, V^{2/3}) ~.
\ee
In both cases we have the universal $2/3$ power, which points us to a synoptic view on the Kraichnan and Kolmogorov scalings.

This explanation of the Kraichnan and Kolmogorov scalings in $2+1$ dimensions has a natural $3+1$ dimensional counterpart.
When we consider the $3+1$ dimensional case, we only have the Yang--Mills CFT, the usual four-dimensional dual to string theory on $\AdS{5}$.
In this case we only have the deformation of the area law and thus the Kolmogorov scaling.
This $\exp(-A^{2/3})$ law would be just the dimensional lift of the same law in $2+1$ dimensions.
As we do not have a membrane theory in $3+1$ dimensions, there would be no counterpart to Kraichnan scaling.

Conversely, the three-dimensional case can be viewed as an $\epsilon$ reduction of the four-dimensional case.
In four dimensions we have quartic potential (which is marginal) and in three dimensions the sixth power, as in the classic result of Wilson and Fisher~\cite{wf}.
When we reduce a turbulently deformed area law from four dimensions to three dimensions, we get either the same result or its volume analogue, where the strings from three-dimensional Yang--Mills thicken into membranes~\cite{tassos}.
The relation between physics in $3+1$ dimensions and $2+1$ dimensions is illustrated in Figure~\ref{fig:one}.
It is on the basis of this that we relate the loop functional associated to turbulence to the Wilson loop in AdS.

\bfig[h]
\centering
\includegraphics[width=1.0\textwidth]{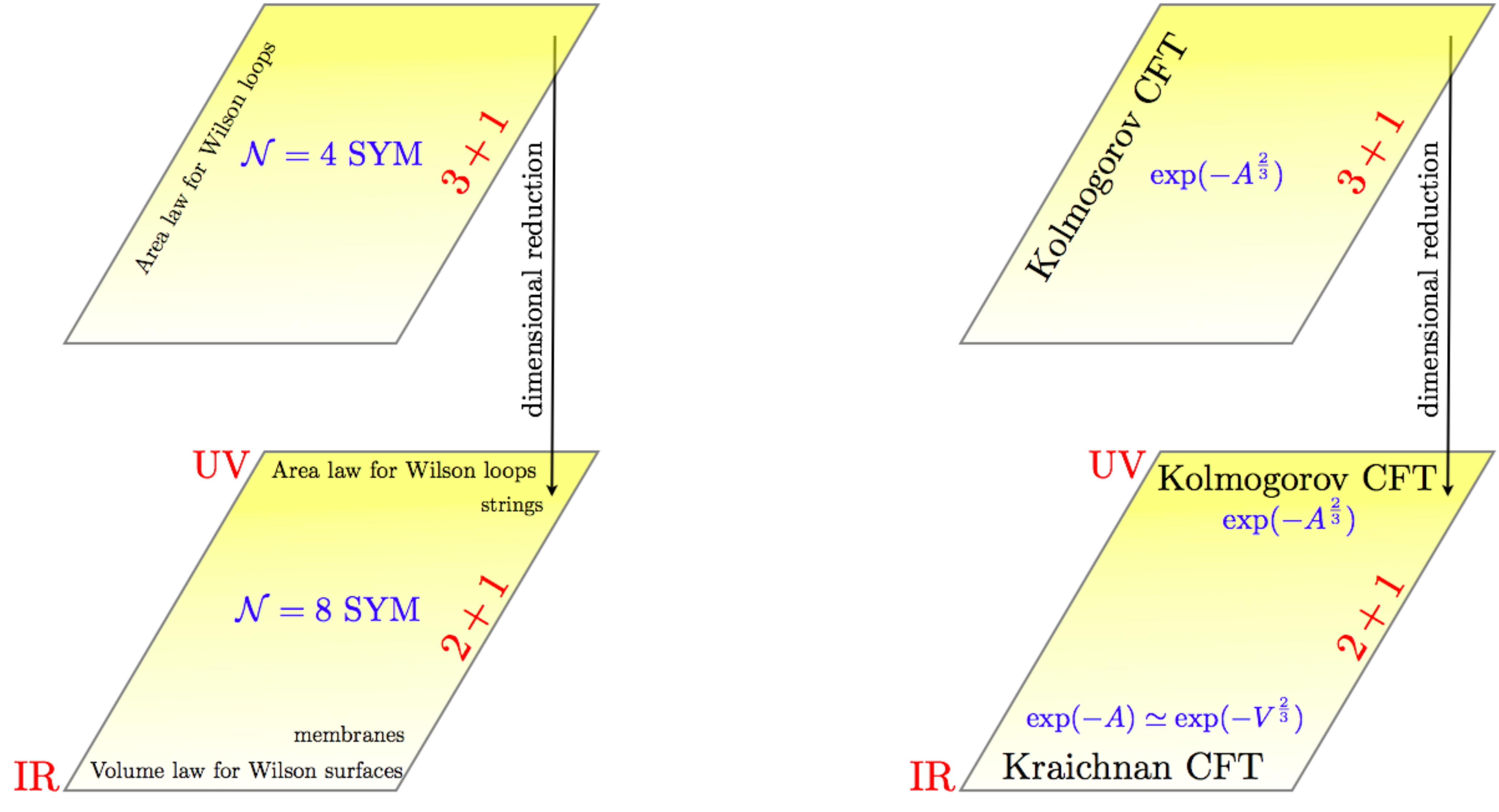}
\caption{
Wilson loops in the ${\cal N}=4$ super-Yang--Mills theory in $3+1$ dimensions exhibit an area law.
In $2+1$ dimensions, the theory reduces to ${\cal N}=8$ super-Yang--Mills.
The area law for Wilson loops (strings) in the ultraviolet flows in the infrared to a volume law for Wilson surfaces (membranes).
Similarly, in $3+1$ dimensions, we have a fluid with Kolmogorov scaling
({\it i.e.}, there is an $\exp(-A^{2/3})$ law for turbulent Wilson loops).
In $2+1$ dimensions, the fluid exhibits the same Kolmogorov scaling in the ultraviolet;
it flows in the infrared to the Kraichnan scaling
({\it i.e.}, there is an $\exp(-A) \sim \exp(-V^{2/3})$ law for turbulent Wilson surfaces).}
\label{fig:one}
\efig

Given the physical picture proposed, the AdS/CFT correspondence elucidates the Kraichnan scaling from a bulk perspective.
By following the membrane to string transition, we see the relation between Kraichnan and Kolmogorov scaling.
Let us explore this point in greater depth.

The expectation value of the turbulent Wilson loop in the Kraichnan scaling regime in $2+1$ dimensions is a standard calculation in AdS/CFT~\cite{maldacena}.
Consider $\AdS{d+1}\times S^p$.
In Poincar\'e coordinates, the metric is
\be
{\rm d}s^2 = g_{\mu\nu}\, dx^\mu\, dx^\nu
= L_{\AdS{d+1}}^2 \left[ \frac{1}{u^2} \left( \frac{-dt^2 + dr^2 + r^2\, d\Omega_{d-2}^2}{L_{\AdS{d+1}}^2} + du^2 \right) + k\ d\Omega_p^2 \right] ~,
\ee
where $d\Omega_p^2$ is the line element on a unit $p$-sphere.
In these coordinates, the boundary is at $u=0$.
In the case of $\AdS{4}\times S^7$, we have $L_{\AdS{4}} = \ell_P (\frac12 \pi^2 N)^{1/6}$, $k = 2$, where $N$ is the flux on $S^7$.
Although $\AdS{4}\times S^7$ is properly an M-theory background, let us examine the stringy Wilson loop.
We restrict to the $\AdS{4}$ factor, which is parametrized by the coordinates $\{t, r, \theta, u\}$, put $\ell_P = \sqrt{\alpha'}$, and Euclideanize the metric by sending $t\mapsto i\, t$.

We consider the embedding $t = {\rm constant}$, $r = r(\sigma)$, $\theta = \tau \in [0,2\pi)$, $u = \sigma\in [0,\infty]$.
The Nambu--Goto action of the string is
\be
S_{NG} = -\frac{1}{2\pi\alpha'} \int d^2\sigma\ \sqrt{\det ( g_{\mu\nu} \pa_a X^\mu \pa_b X^\nu )}
= -\frac{L_{\AdS{4}}}{\alpha'} \int_0^\infty d\sigma\ \frac{r(\sigma)}{\sigma^2}\ \sqrt{1 + \left( \frac{r'(\sigma)}{L_{\AdS{4}}}\right)^2} ~.
\ee
From the Euler--Lagrange equation, we derive as a solution the spacelike Wilson loop with a circular profile of radius $r_0 = L_{\AdS{4}} \sigma_0$ on the boundary:
\be
\frac{r(\sigma)}{L_{\AdS{4}}} = \sqrt{\sigma_0^2-\sigma^2} ~, \qquad \sigma\in [0,\sigma_0] ~.
\ee
The on-shell worldsheet action of the string becomes
\be
S_{NG} = \frac{L_{\AdS{4}}^2\sigma_0}{\alpha'} \left( \frac{1}{\sigma_0} - \frac{1}{\epsilon} \right) ~,
\ee
where we have employed an $\epsilon$-prescription to regulate the integral.
Dropping the divergent piece, we find that the Wilson loop satisfies an area law:
\be
\langle W(A) \rangle = \exp(-S_{NG}) = \exp\left(-\frac{L_{\AdS{4}}^2}{\alpha'}\right) = \exp(-f\, A) ~.
\label{eq:arealaw}
\ee

% {\bf Breaking of susy, keeping only the bosonic part;
% does one have to go to a strongly coupled limit in the bulk as in the story of pure Yang--Mills theory from the bulk point of view?}

We should emphasize several key points before we continue.
Like ${\cal N}=4$ SYM, the membrane theory dual to physics on $\AdS{4}\times S^7$ is a conformal theory.
In particular, as there is no scale within the theory, the expectation value of the Wilson loop is simply a number.
The area law, which is used to signal confinement, or in this case the turbulent phase in the zero viscosity limit of the hydrodynamics, is associated to the Yang--Mills part of the theory.
In order to be explicit about this calculationally, we must heat up the theory;
the temperature then introduces a scale that breaks the conformal invariance.
We use this same picture to assign meaning to the observation that~\eref{eq:arealaw} scales with area.

The background $\AdS{4}\times S^7$ is an M-theory background with Freund--Rubin fluxes~\cite{dnp}.
Via an oxidation procedure~\cite{npstv}, we have $\AdS{4}\times {\mathbb P}^3$ with fluxes as a solution to type IIA string theory.
We can consider the string sigma model within this background.
Introducing a finite temperature black hole in the $\AdS{4}$ bulk generates a scale in the CFT.
Under the assumption of local thermal equilibrium, the physics at long wavelengths is described by fluid dynamics.
Conservation of the energy-momentum tensor in this background yields the Navier--Stokes equations~\cite{minw}.
Following this prescription, in the gauge theory associated to the membrane, the string tension and the viscosity have their natural dimensions.

The turbulent Wilson loop in the Kraichnan regime is the usual Wilson loop associated to the string worldsheet;
the AdS/CFT correspondence enables us to make this identification.
The claim then is that the boundary turbulence in the Kraichnan regime is given by string theory by the Nambu--Goto action in the bulk of $\AdS{4}$.
The same computation of the expectation value of the turbulent Wilson loop in the Kraichnan regime in $2+1$ dimensions should be possible in terms of membrane variables.
We should compute $\exp(-f\, V^{2/3})$ on the boundary via the $\AdS{4}$ bulk evaluation
\be
\exp(-f\, V^{2/3}) = \exp\left[- \left( \int d^3\sigma\ \sqrt{\det(g_{ab}^{\AdS{4}})} \right)^{2/3} \right] ~,
\ee
where once again $g_{ab}$ is the induced metric of the membrane in the $\AdS{4}$ background.
This appears dangerously non-analytic, and it does not have the same asymptotics as the canonical quantum effective actions (the implicit power of $\hbar$ in the denominator is $2/3$), but the scaling law is the same as the area law discussed above, just expressed in terms of unusual variables.
In other words, the bulk on-shell action is the same in both cases.

The physical picture is that the Kraichnan regime of the boundary turbulence is given by the membrane theory in the AdS bulk with an effective non-analytic ``turbulent'' Nambu--Goto worldvolume action $V^{2/3}$.
The two regimes interpolate in three dimensions.
Looking at the behavior of the three-dimensional Yang--Mills theory, in the deep infrared, the turbulent string action may be expressed as a turbulent membrane action.

The inverse renormalization group (RG) is really the holographic RG flow~\cite{rg}, and the Kraichnan scaling in $2+1$ dimensional turbulence is the boundary dual of the string theory in $\AdS{4}$.
The Kolmogorov scaling is related to the Kraichnan scaling in $2+1$ dimensions as the $2+1$ Yang--Mills theory is related to the membrane theory in the deep infrared.
The two Kolmogorov scalings in $2+1$ and $3+1$ dimensions are related by dimensional reductions as the $2+1$ and $3+1$ Yang--Mills theories.

Thus the complete dynamical picture is as follows.
We start with the fluid vortex dynamics.
For a single big vortex we have an effective action given by the Nambu--Goto action.
Now, if this vortex is turned into two, and then four, etc., at the end of the cascade we will have a large number of small vortices.
This is essentially the picture of Kolmogorov.
This means that the area spanned by the vortex ---
not the area of the worldsheet, but the {\em transverse area} whose boundary is given by the vortex ---
is now made of many little vortex areas.
The worldsheet has, from a coarse grained point of view, become effectively a worldvolume!
This is illustrated in Figure~\ref{fig:two}.
In terms of the original Nambu--Goto action for one big vortex we have $\exp(-f\, A) \sim \exp(-f\, V^{2/3})$.

How is this possible?
The idea here is that if turbulence can be formulated as an effective string theory, then we should consider what happens at weak and at strong string coupling.
At strong coupling the turbulent string would turn into a membrane in the same way the usual fundamental string turns into a membrane in M-theory~\cite{dhis}.
If we view the same $\exp(-f\, V^{2/3})$ result from the membrane point of view, then we get our result for the Kraichnan scaling, that is $\exp(-f\, V^{2/3}) \sim \exp(-f\, A)$,
{\it i.e.}, the area law.
The only fact we need to use in this dynamical picture is volume preserving diffeomorphisms as the big vortex decays into many many small vortices.
By lowering the string coupling we get our turbulent string, that is we go from $\exp(-f\, V^{2/3})$ to $\exp(-f\, A^{2/3})$ and this gives the Kolmogorov scaling.

Now conformal symmetry (and AdS/CFT applied to the turbulent string) would tell us that in four dimensions we only have Kolmogorov scaling while in three dimensions we have both Kolmogorov and Kraichnan scaling, in the sense of the flow from ultraviolet to infrared (the inverse cascade).
On top of this we have a natural inverse RG provided by the holographic RG relation between the boundary (where the turbulent string is) and the bulk (where the fundamental string is).

\bfig[h]
\centering
\includegraphics[width=0.8\textwidth]{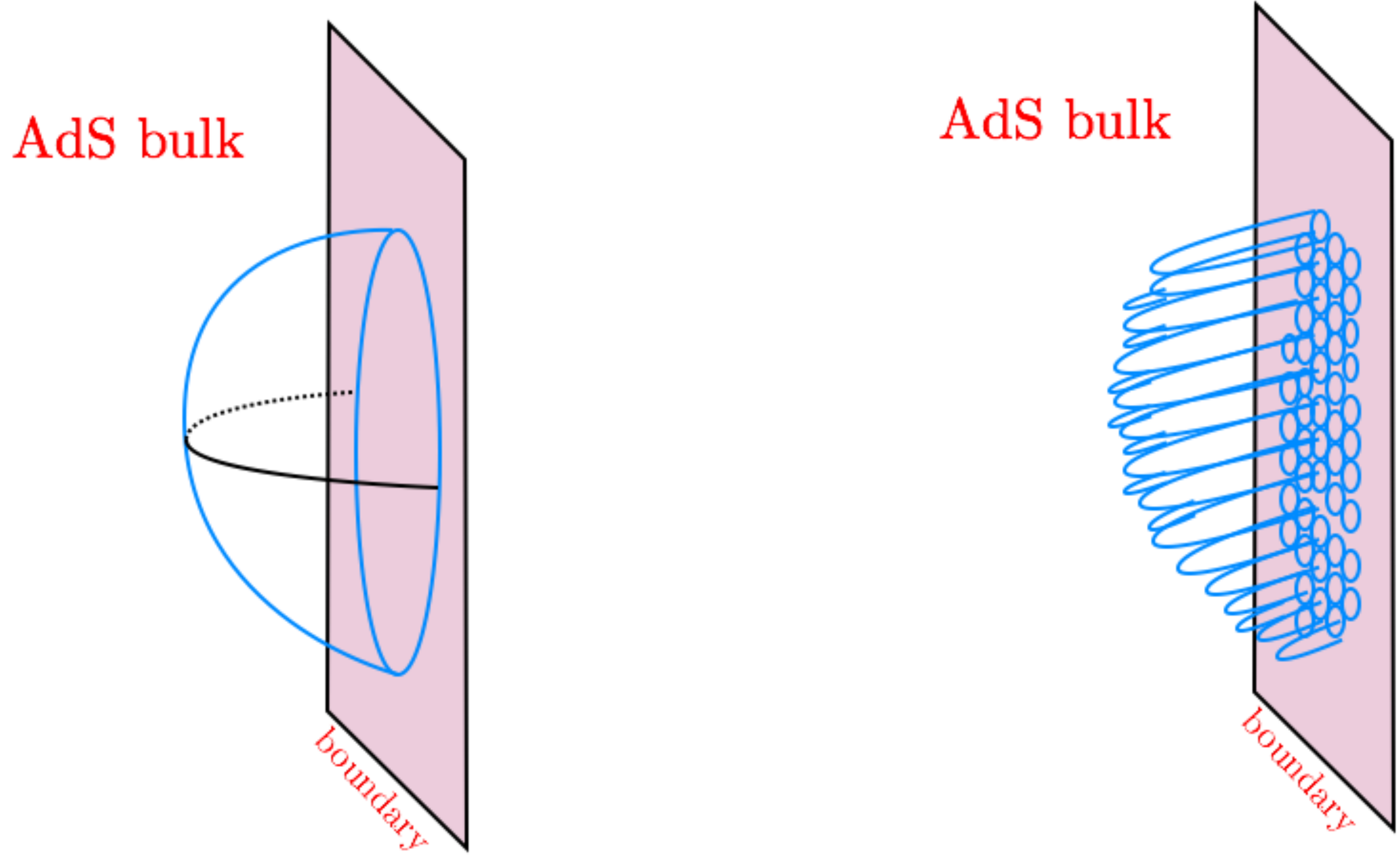}
\caption{
On the left, we have the boundary loop extended into the bulk of AdS as in the computation of the Yang--Mills Wilson loop~\cite{maldacena}.
By comparison, on the right, the boundary loop has broken up in the turbulent regime into many small loops (as in the classic picture of the Kolmogorov cascade).
This boundary picture should be extended, for every little loop, into the bulk of the AdS.
Thus one gets an effective membrane extended into the bulk space.}
\label{fig:two}
\efig

Finally, this picture implies the boundary turbulence/bulk string theory dictionary for the generating functional of velocity correlators as in the AdS/CFT correspondence.
The generating functional of all turbulent correlators of a fluid in the Kraichnan regime in $2+1$ dimensions is given as a bulk string partition function in the semiclassical regime
\be
\left\langle \exp\left(-\int J O(v)\right) \right\rangle = \exp(-S_{{\rm str}}(\Phi)) ~,
\ee
where $O(v)$ are operators constructed out of any power of velocity and its derivatives, $J$ are the sources, and $S_{{\rm str}}$ is the fundamental string action in the appropriate AdS space, which is a functional of the string field in this background (or alternatively the string excitations in this background) which have their boundary values determined by the sources $J$.
The perturbation theory for the Kraichnan scaling is thus organized in terms of natural string variables, as in the usual AdS/CFT dictionary.

The crucial point is that this $2+1$ dictionary has as its ultraviolet completion the Kolmogorov scaling, and this has as its dimensional uplift, the Kolmogorov scaling in $3+1$.
The relationship between the Kolmogorov and Kraichnan scaling is that of gauge theories and membranes in $2+1$ dimensions, or Wilson loops and Wilson surfaces in the two corresponding theories.

In understanding this, it is useful to remember that the interpretation of the Kolmogorov and Kraichnan scalings via the quantum field theory anomaly rests upon the behavior of the three-point functions~\eref{eq:3Kol} and~\eref{eq:3Kr}.
This three-point function in both cases is a pure three-point velocity correlator.
The three-point correlator is divergent, as usual, and has to be regulated in the ultraviolet.
But it is constant if we insert the Kolmogorov and Kraichnan scaling in the infrared.
This agrees very nicely with the picture we have presented.
In the infrared, we have the area law, which is natural for large Wilson loops.
In the ultravolet, the big loop breaks up, as illustrated in Figure~\ref{fig:two}.
The expectation value of the Wilson loop is of course the same.
Thus we can think of this as an anomaly, which has both an ultraviolet and an infrared interpretation.
Rewriting the same area law in terms of volumes and invoking the membrane/string transition, we get the other anomalous behavior, {\it i.e.}, the Kolmogorov law.
The two regimes, Kraichnan and Kolmogorov, are related in our picture, as they should be from the point of view of the underlying three-point velocity correlator.

Note that the boundary turbulent theory is a CFT (and thus similar to~\cite{polyakov}), but its correlator is given in terms of a bulk string theory.
A natural question is to consider the relation (if any) with the conformal fluid explored in~\cite{min}.
Also notice that the fundamental vertex is cubic both from the bulk string field theory and the membrane theory points of view, which is something that has been long expected from the non-linear structure of the Naiver--Stokes equation or its loop counterpart.

A more general lesson of our work may apply to the AdS/CMP correspondence.
One of the major puzzles in the application of AdS/CFT to condensed matter physics~\cite{hh} is why this should even work.
In the case of the gauge/gravity duality we have in mind very well defined physical considerations:
planar diagrams, the large-$N$ expansion, the 't Hooft limit, and the QCD string.
But why should numerous many-body condensed matter systems, which might be governed by various CFTs (classical or quantum), know about string theory and thus gravity?
Our approach to turbulence may provide a clue.
Most of the relevant condensed matter systems currently discussed are quantum fluids (superconductors, superfluids).
For these examples one can write a set of hydrodynamic equations.
These have in fact been written for superfluidity by Landau~\cite{landau}.
Then one can, following Migdal, introduce Wilson loops for these quantum fluids and reformulate the basic equations in terms of new collective variables.
The solution of these equations, {\it i.e.}, the form of the generating functional, is then sought self-consistently, as in our approach.
Viewed from this vantage point the AdS/CFT technology in the context of many-body physics is really an example of a conformal bootstrap in a higher number of dimensions, apart from the usual philosophy that RG is GR, the renormalization group being rewritten in terms of the equations of general relativity.

In sending $\nu\rightarrow 0$, we effectively take a zero temperature limit.
At finite viscosity, a new scale enters, and the fluid mechanics becomes dissipative.\footnote{
For superfluids, while there is no viscosity term, there is a dissipative term with the same scaling as the convective derivative~\cite{v}.
We are grateful to G.~Volovik for a discussion on this point.}
A candidate string dual must exhibit the same property.
To model this explicitly, it may be useful to recall the Caldeira--Leggett setup from condensed matter, in which a system is coupled to a heat bath, which is then integrated out, leading to a dissipative and non-local effective action~\cite{cl}.

In conclusion, we have described a new proposal for a string theory of turbulence.
This proposal explains the Kolmogorov scaling in $3+1$ dimensions and the relationship between the Kraichnan and Kolmogorov scalings in $2+1$ dimensions.
It is natural to speculate that the universal $2/3$ exponent is an indication that one is working in the spacetime foam regime (from a boundary point of view) as suggested in our previous paper~\cite{previous}.
Perhaps this is indicative of the fact that not only can string theory be of use in formulating a theory of turbulence but that the physics of turbulence could provide some guidance to understanding the spacetime foam phase of strong quantum gravity.

\vskip 0.5cm

\noindent
{\bf Acknowledgments:}
We thank Sumit Das, Oleg Lunin, Juan Maldacena, Suresh Nampuri, Leo Pando Zayas, Al Shapere, Steve Thomas, and Grisha Volovik for important discussions on the subject of this letter.
VJ is supported by STFC.
DM is supported in part by the U.S.\ Department of Energy under contract DE-FG05-92ER40677.
YJN is supported in part by the U.S. Department of Energy under contract DE-FG02-06ER41418.
DM and YJN wish to thank Duke University and the organizers of the regional string meeting, Eric Sharpe, Ronen Plesser, and Thomas Mehen, for providing a remarkably stimulating environment for discussion and collaboration.

\end{document}